\documentclass[letterpaper,twocolumn,english,superscriptaddress,letterpaper, nofootinbib]{revtex4}

\usepackage[]{fontenc}
\usepackage[latin9]{inputenc}
\usepackage{textcomp}
\usepackage{amsmath}
\usepackage{graphicx}
\usepackage{amssymb}
\usepackage{bbm}
\usepackage{dsfont}
\usepackage{newtxmath}
\usepackage{dcolumn}
\usepackage{babel}
\usepackage{color}
\usepackage[normalem]{ulem}
\makeatletter

\pdfpageheight\paperheight
\pdfpagewidth\paperwidth

\newcommand{\lyxmathsym}[1]{\ifmmode\begingroup\def\b@ld{bold}
  \text{\ifx\math@version\b@ld\bfseries\fi#1}\endgroup\else#1\fi}

\@ifundefined{textcolor}{}
{%
 \definecolor{BLACK}{gray}{0}
 \definecolor{WHITE}{gray}{1}
 \definecolor{RED}{rgb}{1,0,0}
 \definecolor{GREEN}{rgb}{0,1,0}
 \definecolor{BLUE}{rgb}{0,0,1}
 \definecolor{CYAN}{cmyk}{1,0,0,0}
 \definecolor{MAGENTA}{cmyk}{0,1,0,0}
 \definecolor{YELLOW}{cmyk}{0,0,1,0}
 }

\newcommand{\ys}{YbSb$_{2}$}

\definecolor{dark-red}{rgb}{0.9,0.15,0.15}
\definecolor{dark-blue}{rgb}{0.15,0.15,0.4}
\definecolor{medium-blue}{rgb}{0,0,0.5}

\begin{document}
	
\title{Linear non-saturating magnetoresistance and superconductivity in epitaxial thin films of YbSb$_{2}$ }

\author{Rudra Dhara}
\affiliation{Department of Condensed Matter Physics and Material Science,\\ Tata Institute of Fundamental Research, Mumbai, MH 400005, India}
\author{Pritam Das}
\affiliation{Department of Condensed Matter Physics and Material Science,\\ Tata Institute of Fundamental Research, Mumbai, MH 400005, India}
\author{Sulagna Datta}
\affiliation{Department of Condensed Matter Physics and Material Science,\\ Tata Institute of Fundamental Research, Mumbai, MH 400005, India}
\author{Nilesh Kulkarni}
\affiliation{Department of Condensed Matter Physics and Material Science,\\ Tata Institute of Fundamental Research, Mumbai, MH 400005, India}
\author{Biswarup Satpati}
\affiliation{Surface Physics \& Material Science Division, Saha Institute of Nuclear Physics,\\ A CI of Homi Bhabha National Institute, 1/AF Bidhannagar, Kolkata 700064, India}
\author{Pratap Raychaudhuri}
\affiliation{Department of Condensed Matter Physics and Material Science,\\ Tata Institute of Fundamental Research, Mumbai, MH 400005, India}
\author{Shouvik Chatterjee}
\email[Author to whom correspondence should be addressed: ]{shouvik.chatterjee@tifr.res.in}
\affiliation{Department of Condensed Matter Physics and Material Science,\\ Tata Institute of Fundamental Research, Mumbai, MH 400005, India}

\date{\today}
                           
\begin{abstract}

Rare-earth diantimonides display intriguing ground states often associated with structural order, which can be manipulated in thin film geometries. In this study, we report epitaxial synthesis of one such compound, \ys\/, on III-V substrates using molecular-beam epitaxy. The synthesized thin films exhibit large, non-saturating, linear magnetoresistance across a wide magnetic field range. Additionally, they demonstrate superconducting properties, with a critical temperature of $\approx$ 1.025 K and a critical field of $\approx$ 83.85 Oe, consistent with the reports in bulk single crystals. While \ys\/ has been classified as a Type-I superconductor in its bulk form, our findings provide evidence of a mixed state in the epitaxial thin films. This work paves the way for controlling the electronic ground state in this class of materials through thin film engineering

\end{abstract}

\maketitle     

\section{Introduction}

Rare-earth diantimonides (RSb$_{2}$) exhibit a variety of interesting physical properties, including superconductivity\cite{guo2011dimensional,llanos2024monoclinic,zhao2012type,squire2023superconductivity}, antiferromagnetism and metamagnetism\cite{singha2024anisotropic,trainer2021phase,bud1998anisotropic,zhang2017anisotropic}, heavy fermion behavior\cite{squire2023superconductivity}, charge density waves\cite{luccas2015charge}, and topological nodal lines\cite{guo2011dimensional,o2024ban,qiao2024experimental}. These compounds have a layered structure, where antimony (Sb) atoms form a square net sandwiched between corrugated R-Sb atomic sheets (see Fig. 1(a,b)). The crystal structure of RSb$_{2}$ is highly sensitive to the total electron count on the Sb square net, which is influenced by the valence state of the rare-earth ions. Under ambient pressure rare-earth ions with a 3+ valence typically stabilize an orthorhombic SmSb$_{2}$ structure (space group: Cmca)(see Fig. 1(b)). In contrast, ions with a 2+ valence results in a distorted monoclinic structure, as seen in EuSb$_{2}$ (space group: P2$_{1}$/m). In the case of \ys\/, a small admixture of Yb$^{3+}$ ions at predominantly Yb$^{2+}$ sites stabilizes a more symmetric ZrSi$_{2}$-type orthorhombic crystal structure (space group: Cmcm) (Fig. 1(a)). Similar to other layered intermetallic compounds with two-dimensional square nets, the ground states of these materials depend sensitively on the lattice structure, stacking order, and symmetries. This allows for tunability of their physical properties through external parameters such as pressure and chemical doping\cite{llanos2024monoclinic,squire2023superconductivity}. Epitaxial synthesis of RSb$_{2}$ compounds can open new avenues for controlling their electronic and magnetic properties via substrate-induced bi-axial strain, dimensional confinement, and stabilization of metastable phases.

Here, we report simultaneous observation of large, non-saturating linear magnetoresistance and superconductivity in epitaxial thin films of \ys\/ synthesized on GaSb substrates. Epitaxial integration of rare-earth diantimonides, such as \ys\/, with III-V semiconductors will allow application of substrate induced bi-axial strain, which can be varied due to composition controlled tunability of lattice parameters in III-V substrates, as has been demonstrated for rare-earth monopnictides \cite{inbar2023tuning}. Although epitaxial thin films of \ys\ have been synthesized before\cite{guivarc1993growth}, to the best of our knowledge, there is no report of magnetotransport in \ys\/ or of superconductivity in \ys\/ thin films. Square-net compounds with non-symmorphic space group such as \ys\/ are expected to host linearly dispersive Dirac bands\cite{klemenz2019topological,lee2021topological}, which might play an important role in their magnetotransport properties. Interestingly, magnetoresistance in \ys\/ thin films is found to be non-saturating and remains quasi-linear reaching 350\% at 8.8 T and 2 K.  The non-linear Hall transport in \ys\/ establishes the presence of multiple carriers, which is in accordance with the calculated Fermi surface\cite{sato1999effect}. \ys\/ thin films show superconductivity with a T$_{c}$ $\approx$ 1.025 K with an out-of-plane critical field H$_{c\perp}$  $\approx$ 84 Oe. The estimated superconducting coherence length($\xi$) and penetration depth($\lambda$) in the epitaxial thin films are 198 nm and 404 nm, respectively. Although \ys\/ is reported to be a Type-I superconductor\cite{zhao2012type} in the bulk form, it is a weak Type-II superconductor in the thin film limit, where we find evidence for a mixed state.  

\begin{figure}[htp]
    \centering
    \includegraphics[width=1\columnwidth]{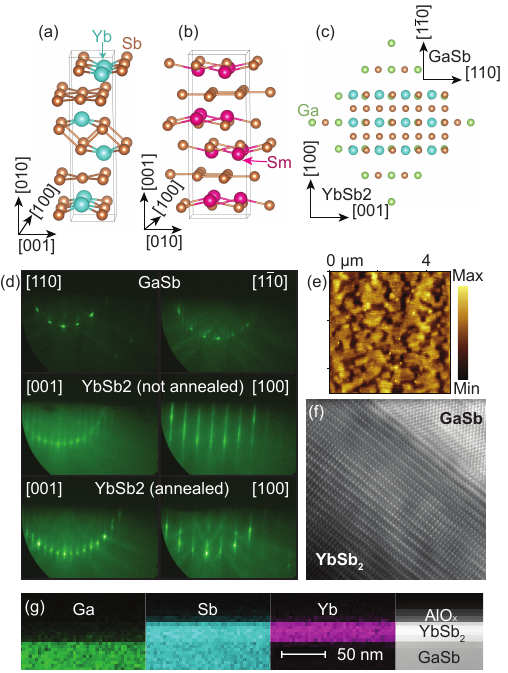}
    \caption{(a) \ys\ and (b) SmSb$_{2}$ crystal structure. (c) Epitaxial relationship between \ys\ and GaSb atomic layers. (d) RHEED images at different stages of the growth of  \ys\/ epitaxial  layers on GaSb. (e) AFM topography of an annealed \ys\/ thin film having an RMS roughness of 0.3 nm over a 5$\mu\/$m$\times$5$\mu\/$m field of view. (f) HR-TEM image showing the atomic arrangement of \ys\ and GaSb with an atomically sharp interface. (g) Cross-sectional EDX scan of the heterostructure showing chemically abrupt interface between GaSb and \ys\/. }
    \label{fig1:structural}
\end{figure}


\section{Experimental Details}
Epitaxial thin films were synthesised inside a molecular-beam epitaxy (MBE) chamber with a base pressure better than 9$\times$10$^{-11}$ mbar. First, around 200 nm of GaSb smoothing layer was grown on a GaSb(001) substrate after the native oxide was thermally desorbed from the substrate under an Sb overpressure. Epitaxial \ys\ atomic layers were synthesised on top of the GaSb homoepitaxial layers with a growth rate of $\approx 0.6$ nm/min, where Yb and Sb were co-deposited from individually calibrated effusion cells. During the growth process \emph{in-situ} reflected high energy electron diffraction (RHEED) was used to monitor the surface evolution. The thin films were capped with $\approx$ 8 nm of AlO$_{x}$ protective layer using an e-beam evaporator before exposing them to the ambient atmosphere, which is essential to prevent sample degradation \cite{chatterjee2019weak, chatterjee2021controlling}. Structural characterization was performed \emph{ex situ} by four-circle x-ray diffraction (XRD) using Cu K$_{\alpha}$ radiation. The atomic scale structure of the \ys\/ thin films was studied by high-resolution transmission electron microscopy (HR-TEM) and energy-dispersive x-ray spectroscopy (EDX) using FEI Tecnai F30 in cross-section. Thin films were further characterized by atomic force microscopy (AFM) and scanning electron microscopy (SEM). Hall bars of width 75 $\mu$m were fabricated using standard optical photolithography technique. Normal state magnetotransport measurements were performed in a commercial PPMS (Quantum Design) using standard low-frequency a.c. lock-in technique. Magnetotransport measurements probing superconductivity were performed inside a home-built $^3\text{He}$ cryostat with a low dc current (5 $\mu\/$A). Penetration depth measurements were performed in a $^3\text{He}$ cryostat using a two-coil mutual inductance technique operating at 30 kHz. In this measurement, a 8 mm diameter superconducting film is sandwiched between a quadrupolar primary coil and a dipolar secondary coil, such that the superconducting film partially shields the magnetic field produced by the primary coil from the secondary. The degree of shielding is dependent on the penetration depth, $\lambda$, and the skin depth, $\delta$ which correspond to the inductive and dissipative response of the thin film, respectively. Therefore, by measuring the out-of-phase and in-phase component of the mutual inductance ($M^{\prime} \quad \& \quad M^{\prime\prime}$) as a function of temperature, we can obtain $\lambda$ and $\delta$ by solving the coupled Maxwell and London equations using finite element analysis. Details of this procedure is given in refs.~\cite{mondal2013phasefluctuationsconventionalswave,turneaure1996numerical,turneaure1998numerical, kamlapure2010measurement, basistha2024low}. For the mutual inductance measurements the amplitude of the a.c. current in the primary coil was kept fixed at 0.01 mA, which corresponds to a peak magnetic field on the film $\approx$ 40 $\mu$Oe. The advantage of this technique is that it allows determination of the absolute value of the penetration depth over the entire temperature range without any prior assumption on the temperature dependence of $\lambda$. 


\section{Results and Discussion}
\ys\ has an orthorhombic crystal structure (\emph{Cmcm}) with bulk lattice parameters, \emph{a} = 4.561 \AA, \emph{b} = 16.72 \AA, \emph{c} = 4.268 \AA \cite{sato1999effect}, where the $a$-$c$ plane share a lattice coincidence with GaSb(001) ($a_{GaSb}/\sqrt{2}$ = 6.095 \text{\AA} /$\sqrt{2}$ $\approx$ 4.31 \text{\AA}) substrates, with a lattice mismatch of \(\sim -5.8\%\) and \(\sim 0.97\%\) along $a$ and $c$ directions, respectively. Accordingly, the atomic layers of \ys\/ are expected to grow with the $b$ axis out-of-plane and the unit cell rotated 45 degrees in-plane with respect to GaSb(001) (see Fig.~1(c)). Furthermore, the surface atomic arrangement of the Sb atoms on the GaSb(001) surface provides an excellent template for epitaxial integration of \ys\/ atomic layers. \ys\/ being the most antimony rich compound in the Yb-Sb system, it can be synthesized under an excess Sb flux. For a Yb:Sb flux ratio of $\approx$ 1:4.8, \ys\/ epitaxial layers can be synthesized between $280^\circ$C and $400^\circ$C. However, at higher temperatures thin film surface becomes rougher. The ideal growth condition is between $310^\circ$C and $330^\circ$C, where thin films with surface rms roughness less than 1 nm can be synthesized without any defect phase, as shown in Figs.~1(e) and 2(a). A reduction in Sb flux under above conditions results in the observation of defect phases corresponding to the different phases belonging to the Yb-Sb system \cite{bodnar1968magnetic, suppl}. RHEED images taken during the growth process is shown in Fig.~ 1(d). Sb-rich c(2$\times$6) surface of GaSb(001) quickly changes to a (2$\times$1) reconstruction on deposition of \ys\/ atomic layers. The film quality in terms of roughness and crystal quality improved greatly when annealed for a short time at $500$\textdegree\/C substrate temperature under an Sb overpressure(see Fig.~1(d)). HR-TEM and cross-sectional EDX images, shown in Figs.~1(f) and 1(g), respectively, confirms sharp hetero-epitaxial interface in the thin films with minimal inter diffusion. Out-of-plane $\theta$-2$\theta$ XRD scan, shown in Fig.~ 2(a), confirms that \ys\/ thin films are single phase with the $b$-axis oriented along the out-of-plane direction. Azimuthal $\phi$ scan, shown in Fig.~2(b), along with the RHEED images (Fig.~1(d)) establish the epitaxial relationship of \ys\/ with respect to GaSb where (010)[001]\ys\/$||$(001)[110]GaSb, in accordance with the earlier report\cite{guivarc1993growth}. The lattice parameters in epitaxial thin films of \ys\/ ($a$ = 4.524 \AA\/, $b$= 16.666 \AA\/, $c$ = 4.308 \AA) are determined from reciprocal space maps (RSM)\cite{suppl}, which are similar to the bulk single crystals. Smooth surfaces in \ys\/ thin films results in the observation of both Laue (Fig.~2(a)) and Kiessig fringes\cite{suppl}, from where we estimate a film thickness of $\approx$ 23 nm. Rocking curve measurements on \ys\/(060) and GaSb(002), shown in Fig.~2(c), establish very high quality of our thin films, where film mosaicity is nearly limited by that of the substrate.  


\begin{figure}[ht]
    \centering
    \includegraphics[width=1\columnwidth]{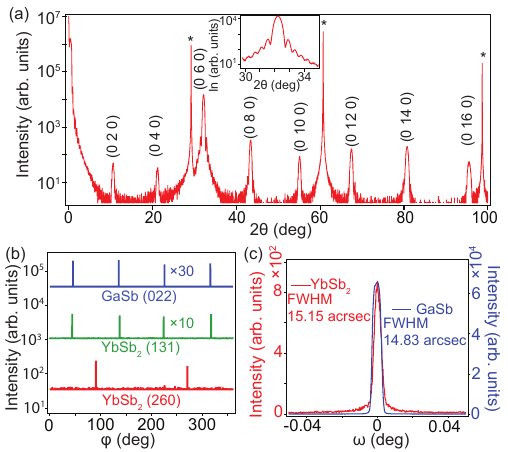}
    \caption{(a) Out-of-plane \(\theta\text{-}2\theta\) scan of a 23 nm thick \ys\ thin film. Substrate peaks are marked by asterisks. The inset shows the Laue oscillation around \ys\/(060) diffraction peak. (b) Azimuthal \(\phi\)-scan of the asymmetric planes of GaSb(022), \ys\/(131) and \ys\/(260). The scans are shifted in the $y$-axis for  a better visualization. (c) A comparison of the rocking curve of \ys\/(006) and GaSb(002) diffraction peaks. }
    \label{fig:XRD}
\end{figure}


Having established the structural aspects of \ys\/ thin films, we now turn towards their magnetotransport properties. All the transport measurements are done on a Hall bar device where the current flows along the [001] direction. At temperatures above 100 K, the electrical transport is dominated by the thermally activated carriers from the narrow-gap semiconducting GaSb substrate. At lower temperatures, metallic behavior is observed  where the transport is dominated by \ys\/ atomic layers and contribution from GaSb substrate is exponentially reduced, as shown in Fig.~3(a)\cite{chatterjee2019weak}. The temperature dependence of resistivity in absence of magnetic field can be well explained by phonon scattering and follows a Bloch-Gr\"{u}neisen (BG) functional form given by $\rho(T) = \rho_{0} + A(\frac{T}{\theta_{D}})^{5}\int^{\theta_{D}/T}_{0} \frac{z^{5}}{(e^{z} - 1)(1-e^{-z})}dz$. The estimated BG temperature is $\approx$ 96 K, which is similar to the Debye temperature of 91.5 K estimated from the Lindemann melting formula in ref.~\cite{bodnar1968magnetic}. However, at lower temperatures, below 15 K, the data is better approximated by $\rho(T) = \rho_{0} + AT^{3}$, which is shown in the inset of Fig.~3(a) (see also Supplementary Information\cite{suppl}). This is indicative of the Bloch-Wilson limit and possibly suggests a dominant role of exchange scattering of conduction electrons off Yb local moments ($s$-$f$ exchange) at low temperatures, as has been observed in other rare-earth intermetallic systems\cite{barnes1967strength}. We note that although a T$^{2}$ dependence of resistivity has been reported below 45 K in the bulk samples\cite{sato1999effect}, a closer inspection in our thin films reveals that the resistivity exhibits a T$^{3}$ dependence below 15 K, above which it exhibits a gradual transition to a linear behavior, as is expected from the low Debye temperature in \ys\/. Therefore, when the fit is performed over a wider temperature range, it erroneously predicts a quasi-T$^{2}$ behavior\cite{suppl}. 


\begin{figure}[ht]
    \centering
    \includegraphics[width=1\columnwidth]{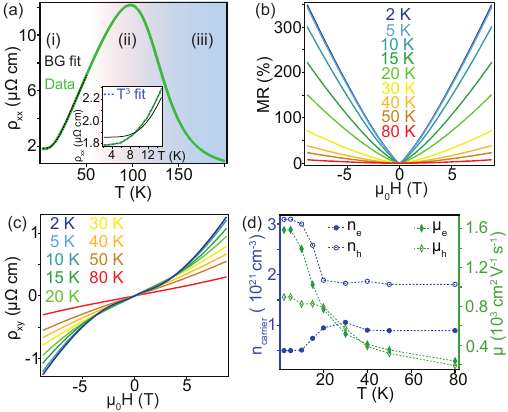}
    \caption{(a) Longitudinal resistivity as a function of temperature. In the regions (i) and (iii) magneto-transport is dominated by the film and substrate respectively. Region (ii) marks the crossover.  Inset shows the comparison between \(\rho(T) = \rho_0 + A T^3\) and BG fit below 15 K. (b) MR as a function of magnetic field at different temperatures. (c) Hall resistivity as a function of magnetic field at different temperatures. (d) Carrier concentration and mobilities estimated from a two-carrier model (see text) at different temperatures.}
    \label{fig3: Transport}
\end{figure}

Magnetoresistance (\(MR = \frac{\rho(H) - \rho(0)}{\rho(0)} \times 100\%\))
in \ys\/ shows non-saturating behavior, which increases dramatically at low temperatures (see Fig.~\ref{fig3: Transport}(b)). At low temperatures, the MR is quite linear and reaches a value of \(\approx 350\%\) at 2 K, while it is only $\approx$ 7\% at 80 K, both at a magnetic field of 8.8 T. Linear, non-saturating magnetoresistance(LMR) in \ys\/ gradually evolves into a more parabolic behavior at elevated temperatures. \ys\/ thin films exhibit a positive non-linear Hall resistivity, shown in Fig.~3(c), which indicates the presence of multiple charge carriers with dominant hole-like conduction. This is in accordance with earlier results on thermoelectric power coefficient in bulk single crystals\cite{bodnar1968magnetic}. The quasi-linear MR behavior at low temperatures indicates that the large MR in \ys\/ cannot be exclusively attributed to electron-hole compensation where the MR is expected to exhibit approximately a quadratic magnetic field dependence\cite{wang2019direct,yuan2016large,sun2022anisotropic}. Accordingly, a two-carrier Drude model given by

\begin{align}
    \rho_{xy}(H) &= \frac{1}{e} \frac{(n_h \mu_h^2 - n_e \mu_e^2) + \mu_h^2 \mu_e^2 (\mu_{0}H)^2 (n_h - n_e)}{(n_h \mu_h + n_e \mu_e)^2 + \mu_h^2 \mu_e^2 (\mu_{0}H)^2 (n_h - n_e)^2} (\mu_{0}H) \label{eq:sigma_xy} \\ \notag
    \\ 
    \rho_{xx}(H) &= \frac{1}{e} \frac{(n_h \mu_h + n_e \mu_e) + (n_e \mu_e \mu_h^2 + n_h \mu_h \mu_e^2) (\mu_{0}H)^2}{(n_h \mu_h + n_e \mu_e)^2 + \mu_h^2 \mu_e^2 (\mu_{0}H)^2 (n_h - n_e)^2} \label{eq:sigma_xx}
\end{align}

cannot simultaneously fit the magnetic field dependence of both longitudinal ($\rho_{xx}$) and Hall ($\rho_{xy}$) resistivities\cite{suppl}. In eqns. 1 and 2, $n_{e}$, $n_{h}$, $\mu_{e}$, $\mu_{h}$ represent electron and hole carrier concentrations, electron and hole carrier mobilities, respectively. In order to obtain an approximate estimate of the carrier concentration and mobilities in \ys\/, a fit to the Hall data was performed using the two-carrier model (eq.\ref{eq:sigma_xy}), where the estimated longitudinal resistivity at zero field ($\rho_{xx}(0)$) from the fit parameters was constrained to be similar to the measured value. The results from such a fit are shown in Fig.~3(d). We obtain hole and electron carrier concentrations at 2 K as $4.95\times10^{20}\ \text{cm}^{-3}$  and $3.1\times10^{21}\ \text{cm}^{-3}$, with the corresponding mobilities of 1586.74 $\text{cm}^2\text{V}^{-1}\text{s}^{-1}$ and 895.63 $\text{cm}^2\text{V}^{-1}\text{s}^{-1}$, respectively. More details can be found in the Supplementary Information\cite{suppl}.  

\begin{figure}[htp]
    \centering
    \includegraphics[width=1\linewidth]{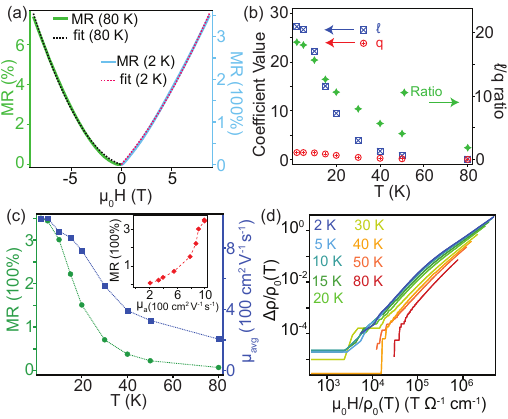}
    \caption{(a) Fits to the MR data at 2 K and 80 K using eqn.~3. (b) Temperature dependence of \textit{l} and \textit{q} coefficients (see text) and their ratio. (c) Evolution of MR and average mobility as a function of temperature. The inset shows MR as a function of average mobility (d) Kohler\textquotesingle\/s scaling of the temperature and field-dependent longitudinal resistivity data ($\rho_{xx}$).}
    \label{fig:enter-label}
\end{figure}
In order to understand the LMR behavior in \ys\/ we adopt a phenomenological model where the observed MR is decomposed into a linear and a quadratic component.
\begin{equation}
\text{MR}(H,T) = l(T)|\mu_{0}H| + q(T)(\mu_{0}H)^{2}
\end{equation}
The temperature dependence of the estimated parameters $l$ and $q$ and their corresponding ratio $l/q$ is shown in Fig.~4(b). While $q$ exhibits a modest increase of 1736.49\% between 80 K and 2 K, increase in $l$ is 17876.25\%, which is larger by an order of magnitude. We note that it is also possible to approximate the MR curves over the entire magnetic field range by a single power law, MR(\%) \(= a' B^{n}\) where $n$ gradually increases with temperature. However, $\rho(H,T)$ in \ys\/ does not follow Kohler\textquotesingle\/s scaling\cite{ziman2001electrons}, as shown in Fig.~4(d). This indicates that the LMR behavior in \ys\/ likely arises due to contributions from different carriers that follow different scaling behaviors under magnetic field, in light of which our attempt at disentangling the linear and quadratic components of MR should be the correct approach. 

 Many different mechanisms have been proposed for the LMR behavior, which has been observed in a variety of material systems. These proposed mechanisms can be broadly classified into two categories viz. those that depend on disorder in the system and the others that rely on the underlying electronic bandstructure. The semi-classical models involving disorder predict the emergence of linear MR both in highly disordered systems\cite{parish2003non,hu2008classical,xu1997large} or in weakly disordered high-mobility samples\cite{wang2014granularity,narayanan2015linear,saini2021linear}. The former scenario can be readily ruled out in the high quality epitaxial thin films of \ys\/ with relatively high carrier mobilities. For the latter case, a transition from a quadratic to a linear behavior is expected at a crossover field ($B_{L}$), where both MR and $B_{L}$ scales with the carrier mobility. No such crossover is observed in our experiments where the quasi-linear field dependence of MR persists till very low fields. Furthermore, it is clearly observed that MR does not scale with the carrier mobility, as shown in Fig.~4(c). LMR behavior has also been predicted to arise from a guiding center diffusion model, which involves smoothly varying disorder potential over much longer length scales compared to the cyclotron period. However, it predicts a Hall angle that is independent of magnetic field and is close to 1, which is not observed in \ys\/ thin films\cite{song2015linear,leahy2018nonsaturating,suppl}. Finally, in all the scenarios discussed above involving extrinsic origin of LMR, such behavior is expected to persist at high temperatures and typically exhibits Kohler\textquotesingle\/s scaling. However, none of these characteristics are observed in \ys\/ thin films. (see Figs.~3(b),4(d)). 

 Among the possible electronic origins of LMR, one possibility is reaching the extreme quantum limit as described by Abrikosov\cite{abrikosov2000quantum} , where the difference between the lowest two Landau levels is comparable to the Fermi energy ($\text{E}_f$). Large carrier concentration of \ys\/, which is of the order of $10^{20} cm^{-3}$, makes this situation extremely unlikely. Another possibility is that the LMR behavior is induced by topological band structure \cite{wang2012linear,feng2015large,zhang2016linear,zhang2022scaling}. A topological nodal line structure has been predicted to exist in a related compound YbBi$_{2}$, where, however, magnetoresistance exhibits a quadratic field dependence ($B^{2}$) and has been attributed to electron-hole compensation\cite{sun2022anisotropic}. The LMR behavior has also been observed in a number of quasi-2D systems with charge density wave (CDW) or spin density wave (SDW) order, which gaps out large portions of the nested Fermi surface\cite{naito1982galvanomagnetic,rotger1994magnetotransport,sinchenko2017linear}. Such a situation can exhibit LMR arising from sharp edges in the Fermi surface, which can dominate MR since it would involve large angular frequency\cite{feng2019linear}. We note that LMR behavior has been observed in a few other square-net rare-earth diantimonides\cite{singha2024anisotropic,fischer2019transport}. Although there is no direct observation of CDW in those systems, it has been speculated to be related to the observed LMR. Although LMR behavior in \ys\/ has many similarities with those observed in density wave systems\cite{feng2019linear}, we have not observed any direct evidence of density wave ordering in \ys\/ thin films. Although our present study establishes that LMR in \ys\/ is likely related to its electronic bandstructure, elucidation of its origin requires further investigation.


\begin{figure}[ht]
    \centering
    \includegraphics[width=1\columnwidth]{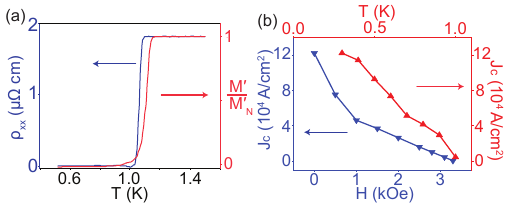}
    \caption{(a) Longitudinal resistivity ($\rho_{xx}$) and the out-of-phase component of mutual inductance ($M^{\prime}$) normalised to its value at 1.4 K ($M^{\prime}_N$),  as a function of temperature at zero magnetic field. (b) The critical current density ($J_c$) as a function of in-plane magnetic field at $T$ = 0.3 K(in blue) and as a function of temperature at $H$ = 0 Oe(in red).}
    \label{fig:fig5_Tc}
\end{figure}



\begin{figure}[ht]
    \centering
    \includegraphics[width=1\columnwidth]{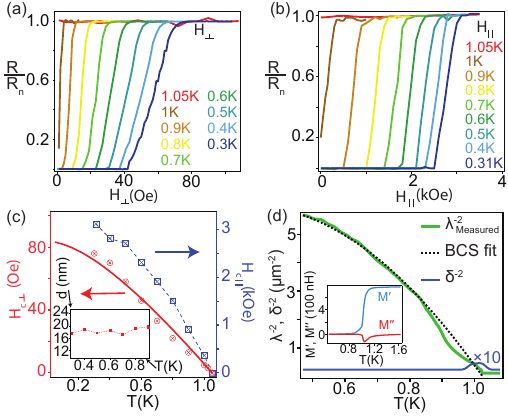}
    \caption{Normalized resistance (R/R$_{n}$) as a function of (a) out-of-plane and (b) in-plane magnetic field at different temperatures. (c) Critical field at different temperatures for in-plane ($H_{c\parallel}$,blue) and out-of-plane ($H_{c\perp}$,red) magnetic field configurations with respect to the thin film plane. The solid red line is the WHH fit of $H_{c\perp}$. The inset shows the calculated thickness from $H_{c\perp}$ and $H_{c\parallel}$ (eqn.~4). (d) Calculated $\lambda^{-2}$ and $\delta^{-2}$ from 2-coil measurements in \ys\/. Corresponding BCS fit is shown with a black dotted line. Inset shows out-of-phase and in-phase components of the mutual inductance from which the $\lambda^{-2}$ is calculated.}
    \label{fig:fig6_sc}
\end{figure}


The epitaxial \ys\/ thin films exhibits superconductivity, shown in Fig.5(a), with a sharp resistive $T_{c}$ of $\approx$ 1.025 K. This is similar to the reported value in bulk single crystals\cite{zhao2012type}. Similar $T_{c}$ is also obtained from two-coil mutual inductance measurement where $M^{\prime} $ shows a sharp drop at $T_{c}$ confirming the bulk nature of the observed superconductivity.  The critical current density at 0.3 K is found to be 1.2$\times$10$^{5}$A/cm$^{2}$ (see Fig.~5(b)). The in-plane ($H_{c\parallel}$) and out-of-plane ($H_{c\perp}$) critical fields were obtained at different temperatures from the magnetoresistance curves below T$_{c}$, shown in Figs.~6(a,b). The critical field ($H_{c}$) was taken to be the value of $H$ where the resistance saturates to normal state resistance. Although \ys\/ has been reported to be a Type-I superconductor in the bulk form\cite{zhao2012type}, sufficiently thin films are expected to exist in a mixed state under a magnetic field when $d$$<\/$$\lambda^{2}$/$\xi$, where $d$, $\lambda$ and $\xi$ are film thickness, penetration depth, and superconducting coherence length, respectively\cite{harper1968mixed,dolan1973critical}. Under such conditions, an estimate of the thin film thickness can be obtained from $H_{c,\perp}$ and $H_{c,\parallel}$ as

\begin{equation}
  d= \sqrt{\frac{6\phi_{0}H_{c\perp}}{\pi\/H_{c\parallel}^{2}}}
\end{equation}

where $\phi_{0}$ is the flux quantum\cite{harper1968mixed}. Following eqn. 4, the estimated superconducting film thickness at all temperatures is $\approx$ $18.4\pm1.4$ nm, shown in Fig.~6(c), slightly smaller than the estimated thickness from x-ray measurements (see Fig.~2(a) and ref.~\cite{suppl}). The temperature dependence of $H_{c,\perp}$ can be well described by a conventional one-band Werthamer-Helfand-Hohenberg(WHH) theory, shown in Fig.~6(c), which describes the orbital limited upper critical field in Type-II superconductors in the dirty limit where spin-paramagnetic effects and spin-orbit scattering can be neglected. The smooth extrapolation of the WHH fit to zero temperature gives $H_{c,\perp}(0)$  = 83.85 Oe, from where the coherence length is estimated as $\xi(0) = \sqrt{\frac{\phi_{0}}{2\pi\/H_{c,\perp}(0)}}$ $\approx$ 198 nm.

In Fig.\ref{fig:fig6_sc}(d) we plot \( \lambda^{-2} \) and \( \delta^{-2} \), obtained from two-coil mutual inductance measurements, as a function of temperature. For an s-wave conventional superconductor in the dirty limit, the temperature dependence of \( \lambda^{-2} \) is given by\cite{tinkham2004introduction,dutta2022superfluid}

\begin{equation}
    \lambda^{-2}(T) = \lambda^{-2}(0) \frac{\Delta(T)}{\Delta(0)} \tanh \left( \frac{\Delta(T)}{2 k_B T} \right)
    \label{eq:pen_depth_BCS_dirty}
\end{equation} 
 
 where $\Delta$ is the superconducting energy gap and $k_{B}$ is the Boltzmann constant. The temperature variation of \( \lambda^{-2} \) can be fitted well to eqn.~5 taking \( \Delta(0) = 1.76 k_B T_c = 0.159 \, \text{meV} \) and assuming the usual Bardeen-Cooper-Schrieffer(BCS) temperature variation of $\Delta(T)$. From the fit, we extract $\lambda(0)$ = 404 nm, which gives a Ginzburg-Landau number $\kappa = \frac{\lambda(0)}{\xi(0)}$ $\approx$ 2.04, consistent with the observed weak Type II superconductivity. Furthermore, the thickness of our thin films is indeed much smaller than $\lambda^{2}$/$\xi$, which makes our estimation of the coherence length consistent, providing evidence for the realization of mixed state in \ys\/ thin films. The larger $\lambda$ and smaller $\xi$ as compared to the earlier reported value in \ys\/ single crystals\cite{zhao2012type} can probably be attributed to larger disorder scattering in our thin film. However, we note that the estimation of $\lambda$ and $\xi$ in ref.~\cite{zhao2012type} involved simplistic assumptions of the underlying band structure in \ys\/, which is complex and anisotropic. In contrast, here we directly estimate $\lambda$ and $\xi$ values without requiring these assumptions.

\section{conclusion}
 In summary, we have investigated epitaxial synthesis of superconducting \ys\ thin films on GaSb(001) substrate and have identified the optimal growth conditions. Our normal state transport revealed large linear, non-saturating magnetoresistance in \ys\/ at low temperatures, which plausibly originates from its complex and possibly topological electronic structure. We confirm the presence of both hole and electron-like carriers in \ys\/, responsible for the observed non-linear Hall effect. The superconducting critical temperature ($T_{c}$) and critical field ($H_{c}$) in \ys\/ thin films are similar to what has been observed in bulk single crystals. Although \ys\/ is reported to be a Type-I superconductor in the bulk form, we provide evidence for the realization of mixed state in epitaxial thin films.  Measured superconducting coherence length ($\xi)$ and penetration depth ($\lambda$) gives a Ginzburg-Landau number ($\kappa$) of $\approx$ 2.04, which establishes \ys\/ thin films as a weak Type-II superconductor. The ability to synthesize these compounds in a thin film form can pave the way to control their physical properties by means of novel tuning parameters such as substrate induced bi-axial strain, proximity effect, and dimensional confinement, available in thin film geometries.

\section*{ACKNOWLEDGEMENTS}

We thank Shikhar Gupta, Devendra Buddhikot, John Jesudasan, and Shankhadip Bhattacharjee for technical assistance. We acknowledge the Department of Science and Technology (DST), SERB grant SRG/2021/000414 and Department of Atomic Energy (DAE) of the Government of India (12-R$\&$D-TFR-5.10-0100) for support. 

\section*{Author Contributions}

 R.D. and S.C. performed thin film growth. N.K. and R.D. performed structural characterization of the thin films.  B.S. prepared thin film lamellae and performed HR-TEM and EDX characterization. R.D. fabricated Hall bar devices and performed magnetotransport measurements in the normal state. P.D. and S.D., under the supervision of P.R., performed magnetotransport and penetration depth measurements down to 300 mK. R.D., P.D., S.D, P.R., and S.C. analyzed the data. R.D., P.R., and S.C. wrote the manuscript. S.C. conceived the project and was responsible for its overall execution. All authors discussed the results and commented on the manuscript.

\section*{Data Availability}

The data that supports the findings of this study are available from the correpsonding author upon reasonable request.

\section*{Competing Interests}
The authors declare no competing interests.

\newpage
\bibliography{bibYS2}

\end{document}